\documentstyle[12pt]{article}
\begin{document}

\title{Debye-Waller factor in He$\rightarrow$Cu(001) collisions revisited: the role of the interaction potentials}
\author{A. \v{S}iber\\
{\em Institute of Physics of the University, 10001 Zagreb, Croatia}\\
B. Gumhalter\thanks{Contact author. Permanent address: Institute of Physics of the University, P.O. Box 304, 10001 Zagreb, Croatia. E-mail: branko@ifs.hr}\\
{\em International Centre for Theoretical Physics, Trieste, Italy
}}

\date{ }
\maketitle

\def\brho{{\hbox{\boldmath$\rho$}}}
\def\bvarepsilon{{\hbox{\boldmath$\varepsilon$}}}
\def\bnu{{\hbox{\boldmath$\nu$}}}
\def\bxi{{\hbox{\boldmath$\xi$}}}
\def\bcalH{{\hbox{\boldmath$\cal{H}$}}}
\def\bcalL{{\hbox{\boldmath$\cal{L}$}}}
\def\bcalW{{\hbox{\boldmath$\cal{W}$}}}

\begin{abstract}
\setlength\baselineskip{2.3ex}
\setlength\textwidth{32pc}

Following the recently accumulated information on the vibrational properties of the Cu(001) surface acquired through single- and multiphonon He atom scattering experiments and the concomitant theoretical investigations, we have reexamined the properties of the Debye-Waller factor (DWF) characteristic of the He$\rightarrow$Cu(001) collisions using the recently developed fully quantal and three-dimensional model of inelastic He atom scattering from surfaces. We have focused our attention on the role which the various He-surface model potentials with their characteristic interaction parameters (range of the interaction, momentum and energy transfer cut-offs etc.) employed in the interpretation of the scattering data may  play in determining the magnitude of the DWF. 
By combining the He-Cu(001) potential whose repulsive and attractive components are both allowed to vibrate with the substrate phonon density of states encompassing anharmonic effects, we obtain the values of the DWF which agree nicely with the experimental data without invoking additional fitting parameters. On the other hand, by taking the phonon  momentum transfer cut-off $Q_{c}$ as an adjustable parameter, as has been frequently exploited in the literature,  all the considered potentials can produce agreement with experiments by varying $Q_{c}$. The magnitudes of such best fit $Q_{c}$ values are compared with those available in the literature and their physical significance is discussed.

\end{abstract}

\newpage
\newcommand{\bq}{\begin{equation}}
\newcommand{\eq}{\end{equation}}

\noindent{\bf 1. Introduction}
\vskip 0.3 cm

A number of experimental studies of the vibronic properties of single crystal surfaces carried out in the past decade
by utilizing thermal energy He atom scattering (HAS)  have emphasized the importance of the microscopic properties of the interactions between He beam atoms and the surfaces investigated. The interpretation of both the single phonon HAS data and the multiphonon scattering spectra required a relatively detailed specification of the form of the interaction potentials and the corresponding matrix elements. The latter proved to be one of the key quantities in establishing a meaningful comparison between experiment and theory \cite{BortoLevi,Cellirev,Santoro,Luo,Benedek,Tommasini}. 

In a number of investigations of the vibrational properties of solid surfaces a successful theoretical interpretation of the relative single-phonon intensities was achieved by using adatom-substrate pairwise potentials and the distorted wave Born approximation (DWBA) for description of the projectile motion in the static component of this potential \cite{BortoLevi,Cellirev,Santoro}. The studies  
of Cu(001) surfaces by HAS emerged as a particularly illustrative example in the context  of the atom-surface interaction potentials. To interpret  the  He$\rightarrow$Cu(001) time of flight (TOF) spectra certain refinements of the theory based on the pairwise potentials \cite{group} have been proposed either through the concept of a pseudocharge model \cite{Luo,Benedek} or the introduction of anisotropic He-substrate atom pair interactions \cite{Tommasini}. In both cases  the modified matrix elements were typified by some cut-off parameters whose variation could significantly affect the one-phonon intensities. However, thus calculated one-phonon scattering spectra could reveal correctly only the relative intensities of distinct phonon modes. 

An additional test of the accuracy of some aspects of the scattering potentials may be provided by the multiphonon spectra whose dependence on the details of the interaction is more complex. Namely, in the multiphonon collision regime  already small changes in the projectile-phonon coupling may give rise to a cumulative effect in the scattering intensity. 
Surprisingly enough, the intensities of the multiphonon scattering  spectra of the He$\rightarrow$Cu(001) collision system \cite{ChemPhys,Hofmann} were relatively successfully reproduced theoretically \cite{Hofmann,comment} by using the earlier expressions for He-surface potentials \cite{group} and two somewhat different approaches to multiphonon HAS. However, no consensus on the values of the characteristic potential parameters has been achieved in these two studies as in one of them \cite{Hofmann} different sets of the interaction parameters have been introduced for single and multiphonon scattering regimes.  

Another basic quantity characteristic of the atom-surface scattering spectra which is sensitive to the features of the atom-surface interaction potentials is the so called Debye-Waller factor (DWF). In the surface scattering experiments it gives a measure of the intensity $I_{0}$ of the elastically scattered specular beam relative to the incoming beam intensity $I_{in}$ and in this respect it differs from the notion introduced in neutron scattering from crystals   \cite{LeviSuhl,dwf}. 
So defined DWF, commonly written in the form $I_{0}/I_{in}=\exp(-2W)$ where $2W$ is the corresponding Debye-Waller exponent, becomes essential in the multiphonon scattering regime because it provides the normalization  proper of the scattering spectrum \cite{BGL,GBL,HAS} which should obey the unitarity principle (optical theorem). Also, in contrast to the measurements of the single phonon scattering spectra, the measurements of the DWF provide values  which  in a sense are "absolute", i.e. unaffected by  the time of flight (TOF) technique, angular and kinematic restrictions, etc.  

A rough estimate of the magnitude of the Debye-Waller attenuation in atom-surface scattering was given long ago by Weare \cite{Weare} and later rederived by Levi and Suhl \cite{LeviSuhl}. They arrived at an approximate expression for the DW exponent as a function of the substrate temperature $T_{s}$ in the form:

\bq
\lim_{T_{s}>\Theta_{D}} 2W(T_{s})=\frac{3(\hbar\Delta k_{z})^{2}}{M_{crys}k_{B}\Theta_{D}}\left(\frac{T_{s}}{\Theta_{D}}\right)=
24\frac{M_{He}E_{i}\cos^{2}\theta_{i}}{M_{crys}k_{B}\Theta_{D}} \left(\frac{T_{s}}{\Theta_{D}}\right).
\label{eq:Weare}
\eq
Here $\Theta_{D}$ is the surface Debye temperature of the substrate, $\Delta k_{z}$ is the change of the projectile momentum normal to the surface, $E_{i}$ and $\theta_{i}$ are the incoming energy and angle of scattering of the projectile, respectively, $M_{crys}$ is the mass of the crystal atoms and $k_{B}$ is the Boltzman constant. $E_{i}$ in this expression is sometimes also corrected for the surface potential well depth $D$ (Beeby's correction \cite{Beeby}) in which case $(\hbar\Delta k_{z})^{2}$ is replaced by $(\hbar\Delta k_{z})^{2}+8M_{He}D$. 
However, the form of the DW exponent (\ref{eq:Weare}) can be justified only in the regime of impulsive scattering \cite{LeviSuhl,dwf} and therefore its validity is of limited range. In particular, for incident energies typical of thermal energy He atom scattering from surface phonons and soft projectile-surface interactions the approximation of impulsive scattering has been shown to become unreliable \cite{HAS} for making quantitative comparisons with the experimental data.

The measurements of the DWF in HAS from metal surfaces have been systematically performed for Ag(111) \cite{Horne}, Pt(111) \cite{Comsa1}, Cu(001) \cite{Lap}, Cu(110) \cite{Comsa2} and Ni(115) \cite{Engel} and theoretical interpretations given within two different approaches developed to treat multiphonon excitations is atom-surface collisions. The calculations of the DWF in HAS from Ag(111) \cite{Modena1} and Pt(111) \cite{Comsa1} were based on a three-dimensional scattering formalism developed in Ref. \cite{CelliMaradudin}. They  reproduced successfully the experimentally observed magnitude and the linear  $T_{s}$-dependence  of the Debye-Waller exponent $2W$ up to  $T_{s}=700 K$  with the "vibrating soft atom model potential" \cite{Modena1}. 
The experimental data on the DWF available for Cu(001) and Cu(110) surfaces were interpreted by carrying out perturbation  expansion of the scattering matrix in a distorted wave basis \cite{ArmandManson}. Although these calculations were essentially one-dimensional as regards the scattered particle dynamics, their real merit lies in  finding that for the atom-phonon coupling to all orders in the lattice displacements the repeated single phonon exchange processes give much larger contribution to the DWF than the simultaneous many-phonon exchange of the same multiplicity \cite{MansonArmand}.
 Upon introducing some adjustable parameters  the DWF was calculated by assuming only the repulsive component of the total potential to vibrate and retaining in the scattering matrix the lowest order dominant contributions in powers of $T_{s}$. Such a truncation of the series for the DWF, which violates the unitarity of the scattering spectrum, gives rise to an artifact in the curvature of the DWF versus $T_{s}$ on a semilogarithmic plot. 

Recently developed fully quantum, three dimensional (as regards the particle dynamics) and unitary multiphonon He atom scattering formalism \cite{HAS} embodies also a general expression for the  DWF as an essential ingredient.  The application of this formalism to multiphonon  He$\rightarrow$Cu(001) collisions  produced a nice agreement of theoretical predictions \cite{comment} with  the experimental multiphonon spectra \cite{Hofmann} by using the same interaction parameters as in the one-phonon theory, i.e. different from those used by the authors of Ref. \cite{Hofmann} to fit the experimental data. 
Hence, the state of affairs regarding the forms of the interaction potentials in HAS from Cu(001) surfaces seems to be critical as different expressions and  parameters have been used to optimally describe the same physical situation.
 
The selection of an adequate model atom-surface potential  to describe inelastic processes in He$\rightarrow$Cu(001) collisions can be also appropriately tested in the evaluation of an "absolute" quantity such as the complete DWF provided the latter is calculated within a genuine  three dimensional scattering model and by taking into account all dominant multiphonon contributions in powers of $T_{s}$. This task is carried out in the present work by making use of the quoted novel multiphonon scattering formalism \cite{BGL,GBL,HAS,comment}. 
In the following sections we focus our attention specifically on the problem of which effects the various model potentials employed in recent interpretations of the one- \cite{Cellirev,Santoro,Luo,Benedek,Tommasini} and multiphonon HAS spectra from Cu(001) \cite{Hofmann,comment} may have on the magnitude of the DWF and its variation with $T_{s}$. 
\vskip 1 cm

{\bf 2. Atom-surface scattering potentials and the DWF}
\vskip 0.5 cm

In a series of earlier publications \cite{BGL,GBL,HAS,comment} we have shown that in  in the case of linear projectile-phonon coupling, which gives by far the largest contribution to the multiphonon processes \cite{MansonArmand}, one can derive a closed form unitary expression for the scattering spectrum valid both in the one- and multiphonon scattering regimes. 
The point of departure in this approach is a distorted wave basis set of projectile wave functions which exactly describe elastic reflections from a flat surface. Inelastic processes are then treated as a perturbation of the distorted waves brought about by the vibrations of the surface. In the weak coupling limit the thus calculated inelastic scattering spectrum reduces, up to a kinematic Jacobian-like factor, to the standard inelastic reflection coefficient calculated in the DWBA \cite{GBL,HAS}. On the other hand, for strong coupling the scattering spectrum is expressed  in terms of uncorrelated and correlated multiphonon scattering processes in which  only one phonon can be exchanged at a certain instant, and the normalization of the entire spectrum is given by the DWF \cite{HAS}.
In the collision regimes typical of HAS experiments the uncorrelated multiphonon processes are dominant over the correlated ones and in this limit the scattering spectrum
acquires a form of the exponentiated  Born approximation (EBA) expression \cite{BGL,GBL,HAS}. The corresponding DWF to all orders in the coupling constant reads \cite{HAS}:

\bq
e^{-2W^{EBA}}=\exp[-\sum_{f\neq i}R_{fi}^{DWBA}],
\label{eq:DWFEBA}
\eq
where $R_{fi}^{DWBA}$ is the temperature dependent one-phonon inelastic reflection coefficient \cite{Cabrera,BortoLevi,Cellirev,Santoro} calculated in the DWBA. $R_{fi}^{DWBA}$ is quadratic in the projectile-phonon coupling and describes the transition of the collision system from the initial state $\mid i \rangle$, characterized by the particle distorted wave quantum numbers ${\bf k_{i}}=({\bf K_{i}}, k_{iz})$ and the initial phonon distribution, to a final state $\mid f \rangle$ in which the particle quantum numbers are ${\bf k_{f}}=({\bf K_{f}},k_{fz})$ and the final phonon distribution differs from the initial one by one phonon quantum. 
Here $\hbar{\bf K}$ denotes the lateral particle momentum and $\hbar k_{z}$ the perpendicular particle momentum outside the range of the static distorting potential $\bar{U}(z)$ of the planar surface. 
Explicitly, one has \cite{HAS,comment}:

\begin{eqnarray}
2W^{EBA}
&=&
\sum_{{\bf Q},j,k_{z}}\left[ \mid {\cal V}^{{\bf K_{i},Q}}_{k_{z},k_{zi},j}(+) \mid^{2}[{\bar n}_{ph}(\hbar\omega_{{\bf Q},j})+1]
+\mid {\cal V}^{{\bf K_{i},Q}}_{k_{z},k_{zi},j}(-) \mid^{2}{\bar n}_{ph}(\hbar\omega_{{\bf Q},j})\right],
\label{eq:WEBA}
\end{eqnarray}
where ${\bf Q}$, $j$ and $\omega_{{\bf Q},j}$ denote the  phonon wavevector parallel to the surface, the branch index and frequency, respectively, and       
${\bar n}_{ph}(\hbar\omega_{{\bf Q},j})$ is the Bose-Einstein distribution of phonons in thermal equilibrium at  $T_{s}$. The on-the-energy-shell one phonon absorption and emission matrix elements are given by \cite{BGL,GBL} 

\bq
{\cal V}^{{\bf K,Q}}_{k_{z},k_{z}',j}(\mp) = 2\pi V^{\bf K,K',Q}_{k_{z},k_{z}',j}\delta_{\bf K',K\pm Q} \delta(E_{{\bf K'},k_{z}'}-E_{{\bf K},k_{z}} \mp \hbar\omega_{{\bf Q},j})
\label{eq:calV}
\eq
where $E_{{\bf K},k_{z}}$ denotes the particle energy. $\mid {\cal V}^{{\bf K,Q}}_{k_{z},k_{z}',j}(\mp)\mid^{2}$  represent first order DWBA probability for a  state-to-state transition $\mid{\bf K},k_{z}\rangle\rightarrow \mid{\bf K'},k_{z}'\rangle$ of the particle. Their seemingly divergent contribution to (\ref{eq:WEBA}) disappears upon conversion of the $\delta$-functions to Kronecker symbols and summation over ${\bf Q}$ and $k_{z}$. 
The matrix elements $V^{\bf K,K',Q}_{k_{z},k_{z}',j}$ of the projectile-phonon interaction $V({\bf r})$ are taken with respect to the distorted waves  corresponding to the projectile motion in  $\bar{U}(z)$. A detailed derivation of these formulae and their application to HAS problems has been given in \cite{HAS,comment,VAS}.

It is evident from Eqs. (\ref{eq:WEBA}) and (\ref{eq:calV}) that the DWF in atom-surface scattering will be sensitive to the form and variations of the dynamical atom-surface potential $V({\bf r})$ through its matrix elements $ V^{\bf K,K',Q}_{k_{z},k_{z}',j}$. These potentials are neither known a priori nor readily  available in analytical form but have to be determined from independent calculations using  various approximate schemes, often yielding them only numerically. 
On the other hand, in the three dimensional numerical calculations  it is often convenient to have analytical representations of both the static and dynamic He atom-surface interactions as this greatly simplifies the computing. This has stimulated the development of the various approximate analytical expressions for the potentials and their matrix elements  as functions of the characteristic interaction parameters such as the strength and the range of the potential, typical cut-offs etc.  

In the majority of theoretical interpretations of inelastic one-phonon HAS experiments on smooth metal surfaces the static projectile-surface interaction is conveniently represented by  a sum of pair potentials $v({\bf r}-{\bf r_{l}})$ acting between the He atom at ${\bf r}$ and substrate atoms at ${\bf r_{l}}$:

\bq
U(\brho,z)=\sum_{\bf l}v({\bf r-r_{l}}),
\label{eq:U}
\eq
and then averaged over the surface \cite{group} to yield the static projectile-surface interaction potential $\bar{U}(z)$.

 The matrix elements of the dynamical interaction in the case of linear atom-phonon coupling  acquire a simple form \cite{Cellirev,Santoro}:

\bq
V_{k_{z},k_{z}',j}^{\bf K,K',Q}=\sum_{{\bf G},\kappa}{\bf u}_{\kappa}({\bf Q},j) {\bf F}_{\kappa}({\bf K-K'},k_{z'},k_{z}) \delta_{\bf K-K',Q+G},
\label{eq:V}
\eq
where the sum ranges over all reciprocal lattice vectors ${\bf G}$ of the surface mesh and the positions ${\bf r}_{\kappa}$ of the crystal atoms in the surface unit cell.  ${\bf u}_{\kappa}({\bf Q},j)$
is the quantized displacement of the crystal atoms corresponding to the  phonon mode characterized by the quantum numbers $({\bf Q},j)$ and the polarization vector ${\bf  e}_{\kappa}({\bf Q},j)$.  The matrix element of the force ${\bf F}_{\kappa}({\bf Q},k_{z'},k_{z})$ exerted on the projectile atom is expressed as \cite{group,Modena3}:

\bq
{\bf F}_{\kappa}({\bf Q},k_{z}',k_{z}) = e^{i{\bf Qr}_{\kappa}} \langle \chi_{k_{z}'}\mid -(i{\bf Q}, \frac{\partial}{\partial z}) v_{vib}({\bf Q},z)\mid \chi_{k_{z}}\rangle,
\label{eq:F}
\eq 
where $\mid \chi_{k_{z}}\rangle$  is the distorted wave describing the particle motion normal to the surface and $v_{vib}({\bf Q},z)$  is a two-dimensional Fourier transform of  the vibrating part $v_{vib}({\bf r})$ of the pair interaction $v({\bf r})$. As yet, there is no general consensus on which part of the total pair potential contributes to $v_{vib}({\bf r})$ and several models have been proposed in the literature. 

Another important parameter characteristic of the various expressions for $v({\bf Q},z)$, and thereby also of $v_{vib}({\bf Q},z)$, is the cut-off wavevector (or wavevectors) $Q_{c}$ which gives an approximate upper bound of the lateral momentum transfer in the one-phonon exchange processes (Hoinkes-Armand effect \cite{Hoinkes}). In the case of exponential potentials of range $1/\beta$ the value of $Q_{c}$ is approximately given by \cite{group,Modena3}:

\bq
Q_{c}=\sqrt{\frac{\beta}{z_{t}}}
\label{eq:Qc}
\eq
where $z_{t}$ is the average value of the He atom turning point in the surface potential $U(\brho,z)$. Numerical evaluation of the matrix elements (\ref{eq:F})  avoids the  explicit introduction of $Q_{c}$ \cite{Tommasini} but the physical effect of the cut off in the space of lateral momentum transfer persists.

Expressions (\ref{eq:DWFEBA})-(\ref{eq:Qc}) provide a framework for a fully three dimensional  calculation  of the DWF and thereby enable a test of the adequacy of the various expressions for the interaction potentials employed in the derivation of the matrix elements (\ref{eq:F}).  

The characteristics of the static He atom-Cu surface potentials have been extensively discussed in the literature \cite{Zaremba,Nordlander,Takada,Toennies,Xenia,Cole}. Quite generally, this interaction is repulsive at short distances and exhibits an attractive van der Waals (vdW) tail at large distances, with a shallow potential well (6-7 meV) whose  minimum is located around 7 bohrs away from the last crystal plane. 
The dynamic He atom-surface potential in the case of linear atom-phonon coupling is obtained as a gradient of the vibrating part of the pair potential, as implicit in Eq. (\ref{eq:F}). However, there is no unanimous  agreement as to which part of the total potential this gradient should involve, which is equivalent to the question whether only the repulsive or both the repulsive and attractive components of the total potential vibrate. 
We shall calculate the force matrix elements (\ref{eq:F}) by using    several different forms of $\bar{U}(z)$ and $v_{vib}({\bf Q},z)$ employed recently in the interpretation of both single and multiphonon scattering data on He$\rightarrow$Cu(001) collisions and then test their relevance by making  a comparison with the experimental values of the DWF.
\vskip 1 cm

{\bf 3. The effect of the potentials on the DWF}
\vskip 0.5 cm

In our assessment of the effects of the various interaction potentials on the DWF pertinent to He$
\rightarrow$Cu(001) collisions we shall investigate all three possibilities of different forms of the static and dynamic interactions. Our point of departure will be the earlier calculated  static He-Cu(001) potential \cite{Xenia}  which was also in a good agreement with empirically determined potentials and experimental fits \cite{Toennies}.
 By requiring that the surface averaged sum of model pair interactions produces as close as possible this potential we may in principle determine the characteristic potential parameters corresponding to $\bar{U}(z)$ and $v_{vib}({\bf Q},z)$ which both derive from $v({\bf r})$. Then, by assuming various forms of $v_{vib}({\bf Q},z)$ (originating either from only the repulsive or both the repulsive and attractive components of $v({\bf r})$)  we may be able to select the physically  relevant $\bar{U}(z)$ and $v_{vib}({\bf Q},z)$ after comparing the calculated and measured values of the DWF.     
To pursue this goal but also remain in correspondence with earlier treatments of single- and multiphonon spectra in obtaining the wavefunctions needed for the calculation of the matrix elements (\ref{eq:F}) we shall fit the $z$ dependence of the potential of Ref. \cite{Xenia} to the following analytical expressions:\\
(i) Exponentially repulsive potential 

\bq
\bar{U}(z)=U_{exp}(z)=U_{0}e^{-\beta z}
\eq
 by requiring that the value and the slope of the two potentials at the He-atom turning points be the same. This implies the pair potential in the form $v({\bf r})=v_{0}e^{-\beta r}$ and  renders $U_{0}$ and $\beta$ as functions of  the incoming energy $E_{i}$ and angle of incidence $\theta_{i}$ of the projectile. This procedure also enables us to estimate the values of $Q_{c}$ using Eq. (\ref{eq:Qc}) because the latter can be most easily justified in the case of exponentially repulsive surface potentials \cite{group}. Here the whole potential is assumed to vibrate, i.e. $v_{vib}(\brho,z)=v(\brho,z)$, and the calculation outlined in Ref. \cite{group} leads to:  

\bq
v_{vib}^{exp}({\bf Q},z)= U_{0}e^{-\beta z}e^{-Q^{2}/2Q_{c}^{2}}
\label{eq:vi}
\eq
 with $Q_{c}$ given by (\ref{eq:Qc}). The nontrivial component of the  interaction matrix elements in (\ref{eq:F}), i.e. $\langle \chi^{exp}_{k_{z}'}\mid -(i{\bf Q}, \frac{\partial}{\partial z}) v_{vib}^{exp}({\bf Q},z)\mid \chi^{exp}_{k_{z}}\rangle$, are then obtainable in analytical form \cite{JM1,Goodman}.     \\
(ii) The Morse potential 

\bq
\bar{U}(z)=U^{M}(z)=U^{M}_{rep}(z)+U^{M}_{att}(z)=D(e^{-2\alpha(z-z_{0})}-2e^{-\alpha(z-z_{0})})
\label{eq:Morse}
\eq
 where the well depth $D$, position $z_{0}$ of its minimum and the range $\alpha$ are determined by requiring that for given $E_{i}$ and $\theta_{i}$ the potentials (\ref{eq:Morse}) and the one computed in Ref. \cite{Xenia} have the same values at the minimum and at the classical turning point. An alternative requirement that the values of the minima and the derivatives at the turning point coincide leads to practically the same fitted Morse  potential. 
The pair potential which leads to (\ref{eq:Morse}) is of the form $v({\bf r})=v_{0}(e^{-2\alpha r}-e^{-\alpha r})$ and the different range of the repulsive and attractive components gives rise to different $Q$-dependence of $v({\bf Q},z)$. 
Here we assume that both components of the pair potential $v({\bf r})$ vibrate which leads to 

\bq
v_{vib}^{M}({\bf Q},z)=D(e^{-2\alpha(z-z_{0})}e^{-Q^{2}/2Q_{c}^{2}} -2e^{-\alpha(z-z_{0})}e^{-Q^{2}/Q_{c}^{2}})
\label{eq:vii}
\eq
 with $Q_{c}=\sqrt{2\alpha/z_{t}}$. 
Here the factor of 2 in the second term in the bracket on the RHS of (\ref{eq:vii}) and the different cutoffs in the $Q$-space ($Q_{c}$ versus $Q_{c}/\sqrt{2}$) arise as a consequence of the different range of the two components of $v({\bf r})$.  Again in this case  the corresponding matrix elements $\langle \chi^{M}_{k_{z}'}\mid -(i{\bf Q}, \frac{\partial}{\partial z}) v_{vib}^{M}({\bf Q},z)\mid \chi^{M}_{k_{z}}\rangle$ are available in analytical form \cite{JM2,Goodman}.\\
(iii) The static potential $\bar{U}(z)$ is given by the Morse potential as in (ii) but only the repulsive component of $v({\bf r})$ is allowed to vibrate. This yields: 

\bq
v_{vib}^{rep}({\bf Q},z)=De^{-2\alpha(z-z_{0})}e^{-Q^{2}/2Q_{c}^{2}},
\label{eq:viii}
\eq
and the corresponding matrix elements  are also expressible in analytical form \cite{JM3,Goodman}.

The interaction potentials (\ref{eq:vi}), (\ref{eq:vii}) and (\ref{eq:viii}) and the corresponding matrix elements should be representative enough to span the range of possible but physically different dynamical  regimes of the projectile-surface encounter which the DWF can be sensitive to. The calculations were performed by using these potentials and the Debye model of surface phonons in Cu parametrized in terms of the surface Debye temperature $\Theta_{D}=290$ K 
which is a mean over the surface projected modes and directions in the  surface Brillouin zone. This effective value is a little higher than the value of 267 K reported in \cite{ChemPhys}  for the Rayleigh wave and the longitudinal resonance in $\langle 100 \rangle$ and $\langle 110\rangle$ directions of the Cu(001) surface, and the value of 280 K found in \cite{Hofmann} as appropriate to the regime of multiphonon scattering from the same surface. 
The difference between the present and the other two temperatures arises because the latter were determined from  fitting the strength of the projectile-surface coupling using a potential in which the effect of the attractive well was neglected.  

Lattice vibrations of copper surfaces exhibit anharmonicity even at moderately high temperatures and in order to take this effect into account we have additionally corrected the   surface phonon density of states for anharmonic effects as discussed in  Refs. \cite{Jayanthi1,Jayanthi2}. The actual effect of anharmonicity on surface phonon dispersion within the two dimensional Brillouin zone pertinent to the Cu(001) surface was estimated from the data presented in Ref. \cite{anhar}. 

In Fig. 1 we show a comparison of the results of calculations of the DWF for the three potentials described above, for $E_{i}$=63 meV and $\theta_{i}=39^{0}$ for which the shadowing effect in scattering from defects should not be very important.  In addition to this we  have also shown for illustration the DWF calculated by using Eq. (\ref{eq:Weare}) with the same $\Theta_{D}$ as in other expressions.
As is seen from Fig. 1, the Morse potential with both components vibrating gives an excellent agreement with  the experimental data. The Morse potential with only the repulsive component vibrating and the adjusted exponentially repulsive potential do not produce  results which would  adequately describe the experiments for the given set of collision parameters, and neither does expression (\ref{eq:Weare}).

Fig. 2 is analogous to Fig. 1 in that it displays the same comparisons, but here for $E_{i}=21$ meV and  $\theta_{i}=31.8^{0}$. The same general trends as observed in Fig. 1 persist for this set of collision parameters as well although the agreement is not as good as at $E_{i}=63$ meV. Presumably this is due to the presence of diffuse elastic scattering which is more pronounced at lower collision energies but hasn't been accounted for by our model. A relatively good agreement between the measured DWF values and those calculated from Eq. (\ref{eq:Weare}) is a mere coincidence which doesn't occur at other collision energies and scattering angles (c.f. Fig. 1). 

A poor description of the DWF in terms of the adjusted exponential potential (designated by (\ref{eq:vi}) above) in Figs. 1 and 2  signifies the importance of the presence of the potential well, and this becomes more apparent as the normal component of the projectile incoming energy is lowered. 
On the other hand, the large difference between the results for the same Morse potential but with different vibrating components arises from smaller interaction matrix  elements in the case of the total vibrating potential. 
This is mainly because the derivative of the complete  potential changes sign at the well bottom giving rise to cancellation effects in the scattering amplitude (\ref{eq:F}). Another point worth emphasizing is that the lateral momentum cut-off function associated with the attractive component of the total vibrating potential attenuates  much faster than the one associated with the repulsive vibrating component, giving correspondingly smaller contribution to (\ref{eq:F}). This simply reflects the fact that the  potentials of longer range give rise to smaller inelastic contributions for given energy and momentum transfer. 
The same trends regarding  the three model potentials are also retrieved for other scattering angles at He incoming energies of 63 and 21 meV. 
Hence, in the scattering regime characterized by the present collision parameters all these findings render, out of the potentials (\ref{eq:vi}), (\ref{eq:vii}) and (\ref{eq:viii}), the total vibrating potential designated by (\ref{eq:vii}) as the best model potential underlying the physics of the DWF in the present collision system.

In Figs. 3 and 4 we present further comparisons of the available  experimental values of the DWF \cite{Lap} with  the  ones calculated using the potential (\ref{eq:vii}). Here we want to reiterate that in these calculations we haven't introduced any free parameters which could be adjusted so as to obtain a good  fit to the experimental data but that all the parameters typical of the potential and the phonon density of states have been either predetermined or obtained within the present model.

Fig. 3 displays the results of the calculations for incident energy $E_{i}=63$ meV and the angles of incidence used in the experiments \cite{Lap}. The agreement between the calculated and experimental results is very good in the whole range of experimental $\theta_{i}$ in which also the earlier empirical fits were successful \cite{Lap}. An  exception occurs only at $\theta_{i}=71.6^{0}$  for which due to the low normal component of the incident energy  the contribution to scattering by surface defects (steps, kinks, adatoms, vacancies etc.) may become important. 
In this situation two effects come into play. 
First, the diffuse scattering by static defects reduces the DWF at incident angles nearer to grazing because of the shadowing effect and, second, the enhanced anharmonicity of vibrations associated with defects would further suppress the magnitude of such DWF. 

The general trend of reduction of the DWF with the angle of incidence closer to the normal signifies the strongest coupling of the He atom to perpendicular surface vibrations. A deviation of the experimental results from this general behaviour for $\theta_{i}=19^{0}$ and 400 K$\leq T_{s} 
 \leq$800 K is probably an artifact connected with the early measurements \cite{Lap}. 

Fig. 4 displays the experimental and theoretical DWF values for incident energy $E_{i}=21$ meV. Although the trends observed here are the same as in Fig. 3, due to the lower incoming energy the deviations between the two sets of data becomes apparent  even at lower incoming angle. For $\theta_{i}=73.5^{0}$ the corresponding $E_{i}^{z}$ is already so small (1.7 meV) that the scattering from  defects may give rise to contributions to the Debye-Waller exponent $2W$ which are larger than that induced by phonons. As our model doesn't encompass this type of effects, the difference between the experimental and calculated values of the DWF for such collision parameters is not surprising.     

A crucial element in obtaining a good agreement between the calculated and experimental values of the DWF without invoking the fitting parameters was a realistic form of the He-surface interaction potential with both components allowed to vibrate, and the fact that the attractive component is much more sharply cut off in the $Q$-space due to its longer range. Thereby the attractive component gives only a correction to the one-phonon matrix elements (\ref{eq:F}) whose magnitude is dominantly determined by the repulsive component of the interaction. 

Quite generally, the matrix elements (\ref{eq:F}) are so sensitive to even small variations of  $Q_{c}$ that the latter may be taken as a fit parameter which could be adjusted so as to produce a good agreement between the calculated and measured values of the DWF for any of the potentials discussed under (i)-(iii) above. Fig. 5 shows a comparison of the magnitudes of such best fit $Q_{c}$ values together with the values available in the literature. However, a physical justification of such best fit $Q_{c}$'s may not be easy in all the cases considered because some of them considerably deviate from the values predicted by the expression $\sqrt{2\alpha/z_{t}}$ here found to provide a consistent description of the experimental results for the DWF in He$\rightarrow$Cu(001) collisions.  

\vskip 1 cm

{\bf 4. Conclusions}
\vskip 0.5 cm

In this work we have studied the effect of the characteristics of the various model interaction potentials on the magnitude of the Debye-Waller factor (DWF) in He$\rightarrow$Cu(001) scattering. For this collision system the experimental data on the DWF \cite{Lap}, single phonon \cite{Luo,Benedek,Tommasini} and multiphonon spectra \cite{ChemPhys,Hofmann} are available for a wide range of collision parameters (incident energy and angle) and surface temperature which all facilitates a comparison between the theoretical predictions and experimental results. 
In our theoretical description we have employed the earlier developed realistic, fully three-dimensional quantum scattering model to calculate the DWF to all orders in the projectile-phonon coupling \cite{HAS}, here assumed to be linear in phonon displacements. 
The calculations were carried out within the so called exponentiated Born approximation (EBA) which takes into account uncorrelated  multiple phonon exchange processes which have been shown earlier to give a dominant contribution to the scattering spectra in the collision regime studied \cite{BGL,GBL,HAS}. 
By considering several types of He-surface interaction potentials we have shown that the model potential which correctly reproduces the gross features of the earlier calculated static He-Cu(001) interaction \cite{Xenia} and whose both components, repulsive and attractive, are assumed to vibrate, produces results which nicely agree with the experimental data without invoking any fitting parameters. Important elements in obtaining this agreement were the difference in the cut off wavevector $Q_{c}$ characterizing the longer range attractive and shorter range repulsive components of the vibrating potential and the variation of the phonon dispersion with temperature due to the anharmonicity of surface vibrations. 
Other model potentials which have also been frequently used in the interpretation of the DWF and the single phonon data, like the vibrating exponentially repulsive potential, fail to reproduce the experimental values of the DWF in the  He$\rightarrow$Cu(001) collision system. This is mainly because they do not take into account the attractive component of either the static or dynamic potentials which become important at lower incoming energies. On the other hand, at very low normal incoming energies the scattering from defects starts to affect the magnitude of the DWF. In this limit the agreement between the measured values  and those predicted by the present model which doesn't account for scattering by defects, is lost.

We have further shown that it is in principle possible to reproduce the measured DWF data with all the model potentials considered by assuming the characteristic cut off wavevector as an adjustable parameter for each set of collision parameters. Thereby one can determine the best fit $Q_{c}$ values as function of the normal component of the projectile incident energy for the studied model potentials and correlate them with the values given in the literature \cite{Luo,Benedek,ChemPhys,Hofmann,comment}. 
The best fit $Q_{c}$ values for the potential with both vibrating components were found to be practically identical to those of our no-fit-calculations, thereby confirming the  consistence of this potential and the validity of the results obtained by using it. They are also found to be very close to the majority of the values cited in the literature in connection with the theoretical interpretation of both single- and multiphonon scattering spectra which gives additional credibility to these and the present calculations.    

\vskip 1 cm

{\bf Acknowledgments:}

One of the authors (B.G.) would like to acknowledge the Senior Associateship Award which enabled his stay at the International Centre for Theoretical Physics in Trieste. The work in Zagreb has been supported in part by the National Science Foundation grant JF-133.

\newpage

\listoffigures
\vskip 1 cm

\noindent{\bf Fig. 1.}  Debye-Waller factor as function of the substrate temperature for He incoming energy $E_{i}=63$ meV and incident angle $\theta_{i}=39^{0}$ calculated with three different interaction potentials. Full squares denote experimental datapoints \cite{Lap} and the long dashed, full and dashed-dotted lines denote the values calculated with the adjusted exponentially repulsive potential (\ref{eq:vi}), total vibrating Morse potential (\ref{eq:vii}) and repulsive vibrating Morse potential (\ref{eq:viii}), respectively. The short dashed line denotes the DWF calculated from Eq. (\ref{eq:Weare}). 
\vskip 0.5 cm

\noindent{\bf Fig. 2.} Same as in Fig. 1 but for $E_{i}=21$ meV and $\theta_{i}=31.8^{0}$.
\vskip 0.5 cm

\noindent{\bf  Fig. 3.} Comparison of calculated (full lines) and experimental values of the DWF for $E_{i}=63$ meV and angles of incidence 19$^{0}$ (squares), 39$^{0}$ (circles), 51$^{0}$ (triangles), 61.7$^{0}$ (inverted triangles) and 71$^{0}$ (diamonds). 
\vskip 0.5 cm

\noindent{Fig. 4.} Same as in Fig. 3 for $E_{i}$=21 meV and incident angles  31.8$^{0}$ (squares), 55.5$^{0}$ (circles) and 73.5$^{0}$ (triangles). 
\vskip 0.5 cm

\noindent{\bf  Fig. 5.} Values of the cut-off wavevector $Q_{c}$ as function of the normal component of He atom incident energy as determined from various procedures. Dashed line:  $Q_{c}=(2\alpha/z_{t})^{1/2}$ where $\alpha$ is the range parameter of the Morse potential. This form of $Q_{c}$ has been used in the calculations leading to Figs. 3 and 4. Full squares: best fit $Q_{c}$ values for the vibrating repulsive component of the Morse potential. Full triangles: best fit $Q_{c}$ values for the adjusted exponential potential. Full circles denote the best fit $Q_{c}$ values for the vibrating Morse potential and they practically coincide with the dashed line for $E_{i}^{z}\geq 15$ meV.  Open symbols denote  $Q_{c}$ values pertinent to exponentially repulsive potentials given in the literature: open squares are from \cite{ChemPhys}, open circle from \cite{comment} and open triangle from \cite{Hofmann}.

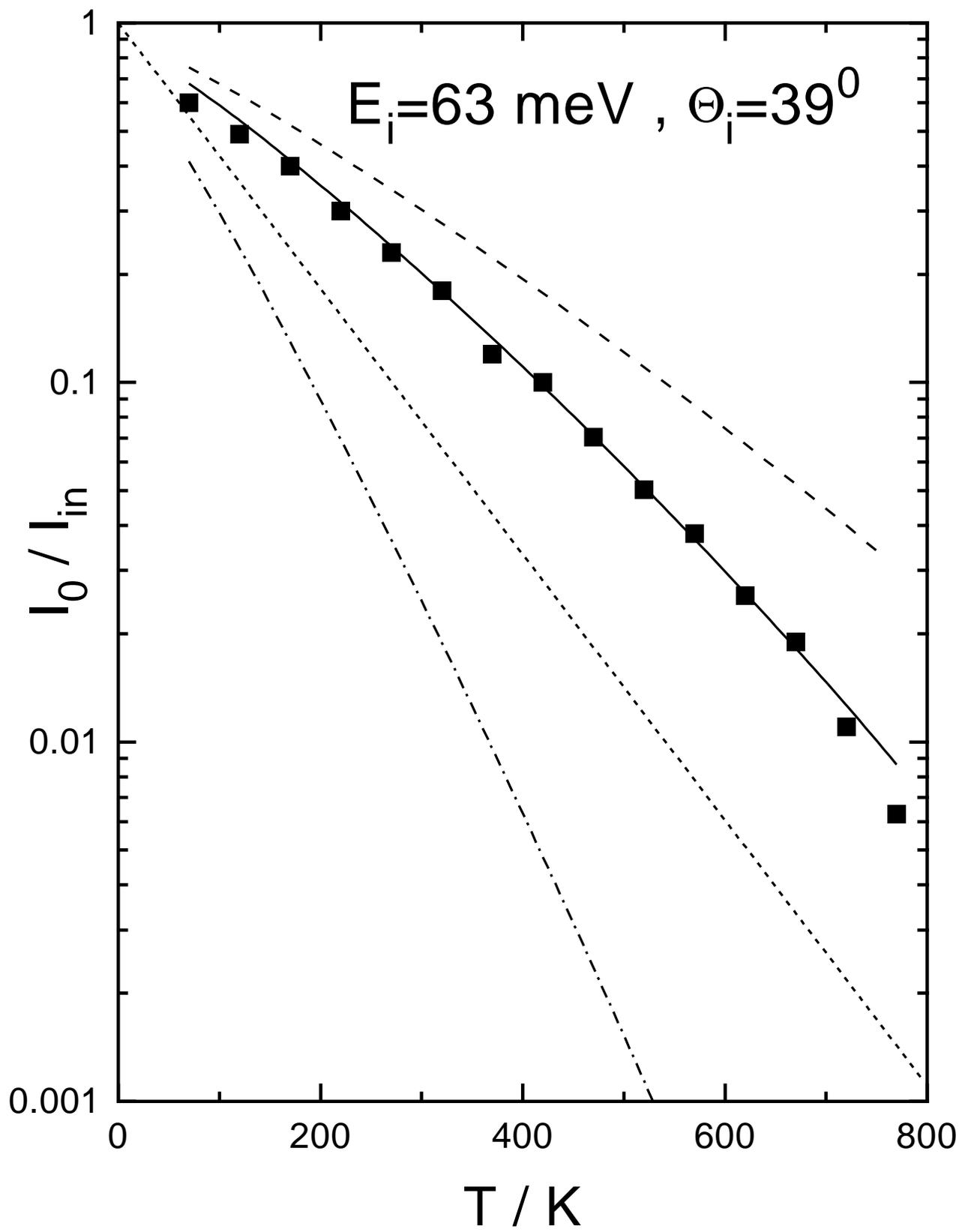
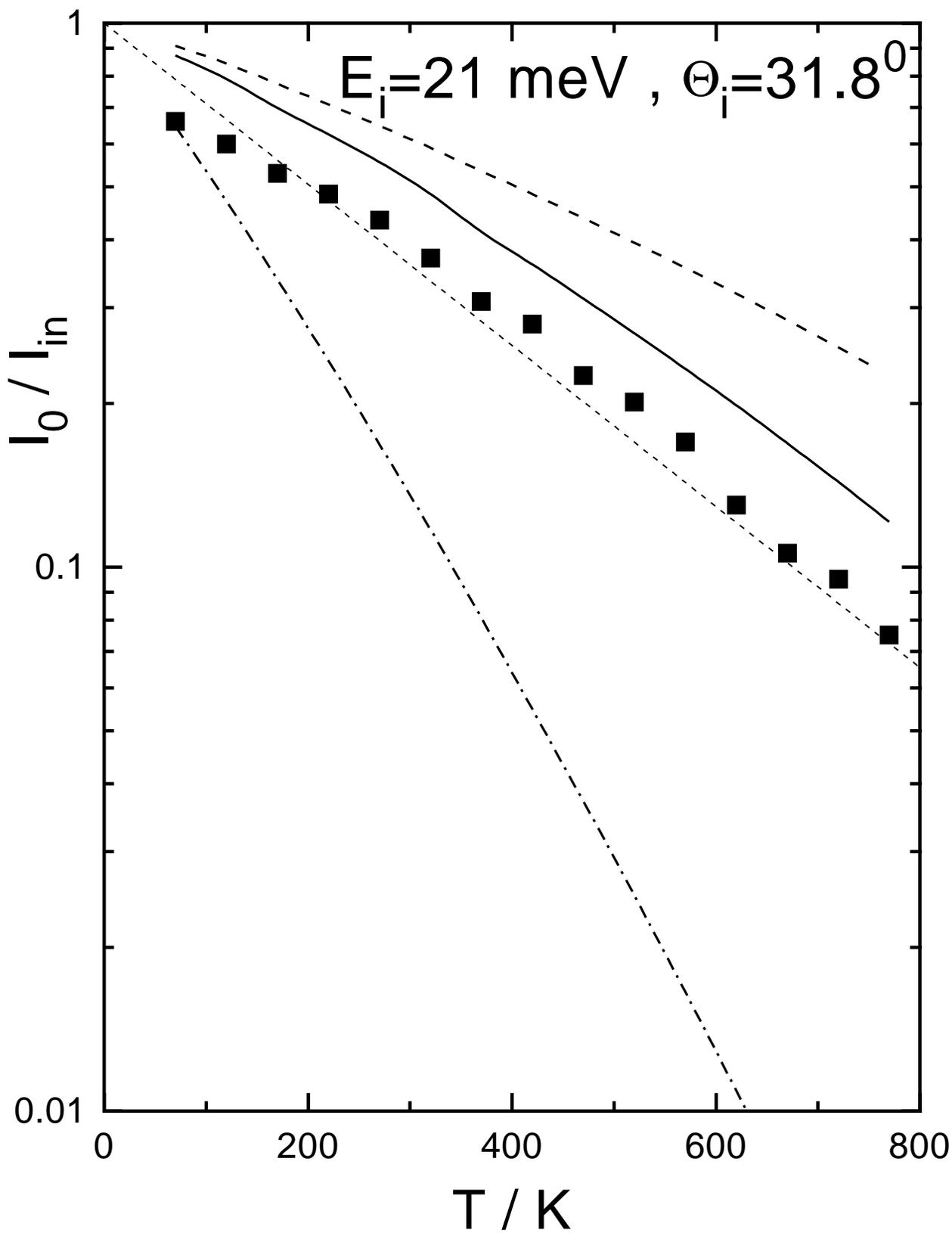
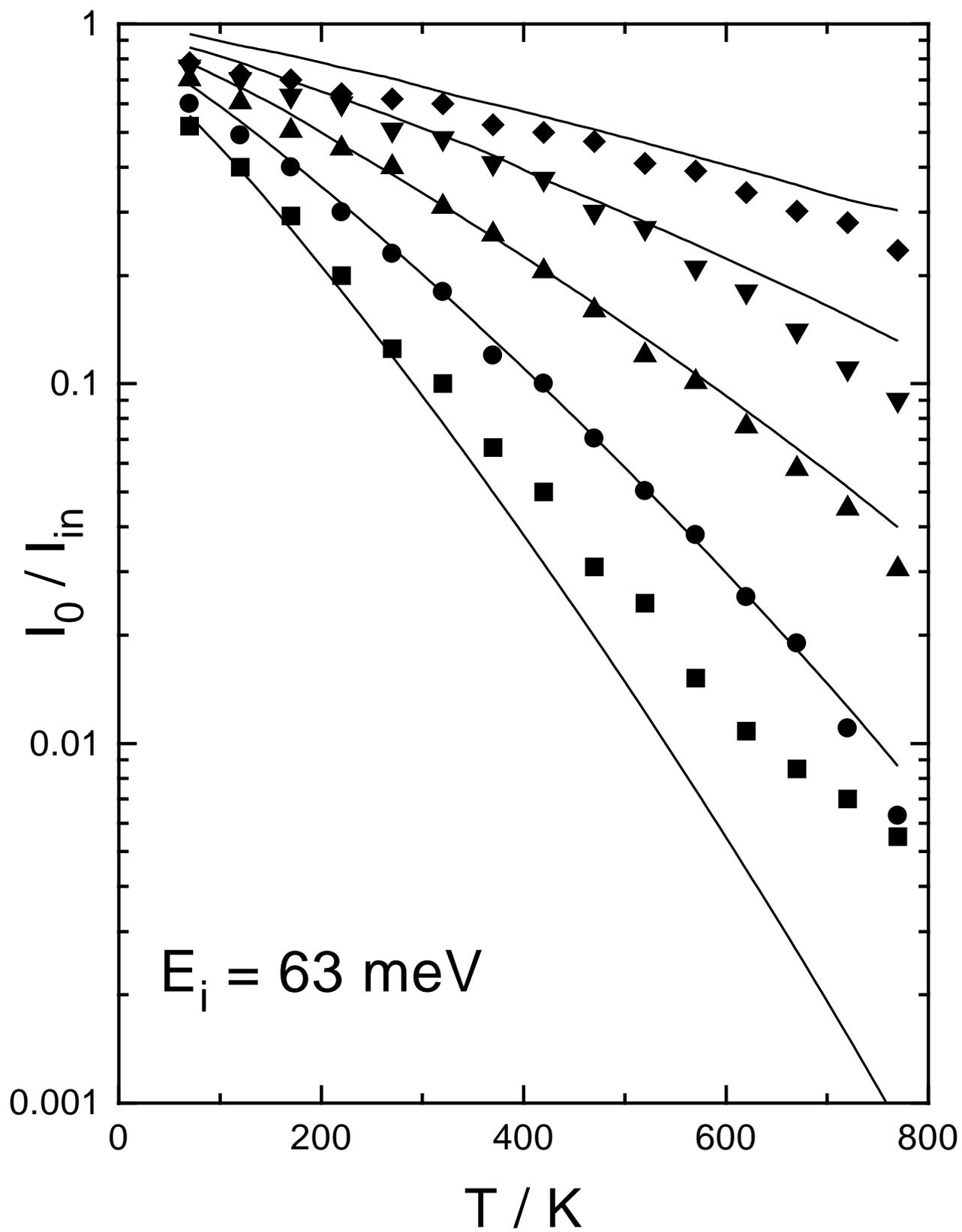
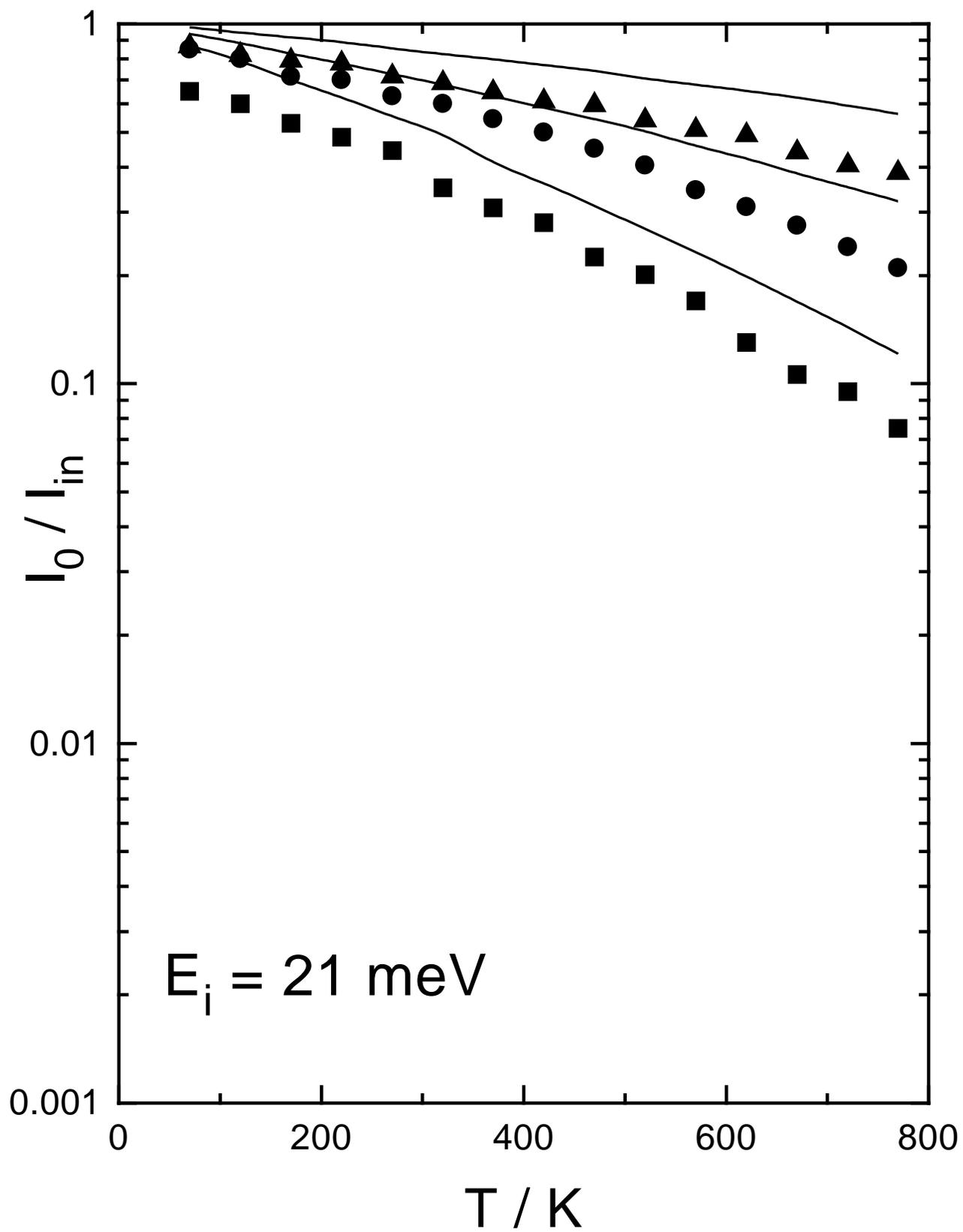
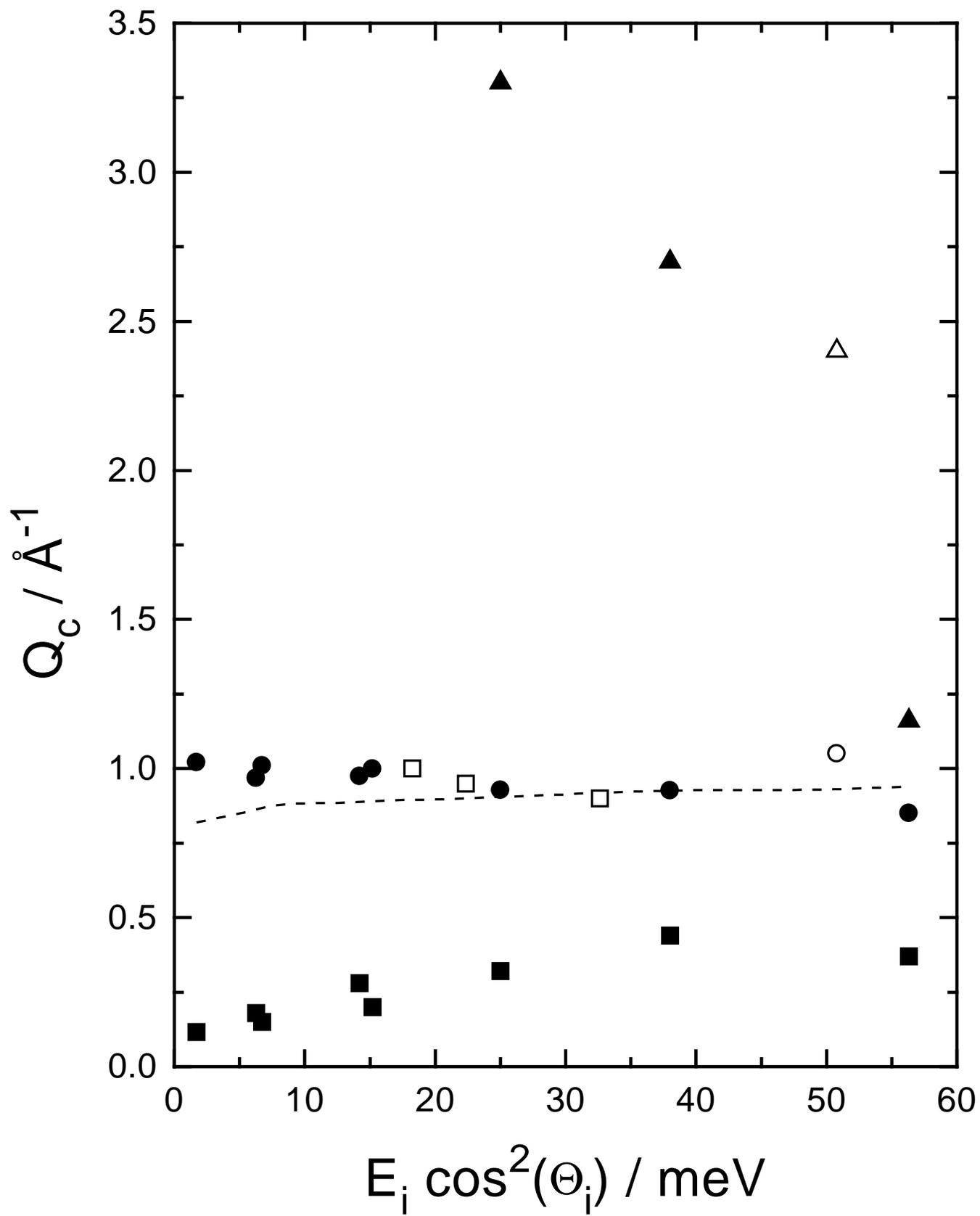
\end{document}